\documentclass[twocolumn,showpacs,preprintnumbers,amsmath,amssymb,showkeys,floatfix]{revtex4}

\usepackage{graphicx}
\usepackage{dcolumn}
\usepackage{bm}

\newcommand{\trn}{^{\scriptscriptstyle \top}} 

\begin{document}

\title{Speed-gradient principle for nonstationary processes in
thermodynamics}

\author{Alexander L. Fradkov}
  \email{alf@control.ipme.ru}
  \affiliation{Institute for Problems of Mechanical Engineering,
Russian Academy of Sciences, \\ 61, Bolshoy ave. V.O., 199178,
Saint Petersburg, Russia}

\date{\today}

\begin{abstract}
The  speed-gradient variational principle (SG-principle) is
formulated and applied to thermodynamical systems. It is shown
that Prigogine's principle of minimum entropy production and
Onsager's symmetry relations can be interpreted in terms of the
SG-principle and, therefore, are equivalent to each other. In both
cases entropy of the system plays a role of the goal functional.
The speed-gradient formulation of thermodynamic principles provide
their extended versions, describing transient dynamics of
nonstationary systems far from equilibrium. As an example a model
of transient (relaxation) dynamics for maximum entropy principle
is derived.
\end{abstract}


\maketitle

\section{Introduction}

The equations of motion for physical   systems are often derived
from variational principles: principle of least action, maximum
entropy principle, etc. \cite{Lanczos64,Gyarmati70,Van95}. In
thermodynamics two of such principles have become well known
during last century: Prigogine's principle of minimum entropy
production and Onsager's symmetry principle for kinetic
coefficients. Authors of both results were awarded with Nobel
prizes. Variational principles are based on specification of a
functional (usually, integral functional) and determination of
real motions  as points in an appropriate functional space
providing extrema of the specified functional.

In addition to integral principles,  differential (local) ones
were proposed: Gauss principle of least constraint, principle of
minimum energy dissipation and others. It has been pointed out by
M.~Planck \cite{Plank14} that the local principles have some
preference with respect to integral ones because they do not fix
dependence of the current states and motions of the system on its
later states and motions.  In \cite{Fra90,F03,F05} a new local
evolution principle, so called {\bf speed-gradient (SG) principle}
originated from the SG-design principle of nonlinear control
theory \cite{Fra90,FP98} was proposed and illustrated by a number
of examples from mechanics. In \cite{F07} SG-principle was
extended to the case of systems with constraints.

 This paper is devoted to application of the
SG-principle to thermodynamics. First, the formulation of the
SG-principle is recalled. Then it is shown that Prigogine's and
Onsager's principles can be interpreted in terms of the
SG-principle and, therefore, are equivalent to each other. In both
cases entropy of the system plays a role of the goal functional.
The speed-gradient formulation of thermodynamic principles provide
their extended versions, suitable for the systems far from
equilibrium. Moreover, it may describe their nonstationary,
transient dynamics. In the paper SG-principle is applied to
derivation of  transient (relaxation) dynamics for a system driven
by maximum entropy principle.

\section{Speed-gradient variational principle}

Consider a class of physical systems described by systems of
differential equations
\begin{equation}
\label{a21} \dot x=f(x,u,t),
\end{equation}
where $x$ is $n$-dimensional vector of the system state, $u$ is
$m$-dimensional vector of free (input) variables, $\dot x =
dx/dt,~t\ge 0$. The problem of modelling system dynamics can be
posed as the search of a law of changing $u(t)$ in order to meet
some criterion of ``natural", or ``reasonable" behavior of the
system. Let such a behavior be specified as a tendency to achieve
a goal, specified as decreasing the value of the {\it goal
functional} $Q(x)$, where $Q(x)$ is given {\it apriori}.  The
first step of the speed-gradient procedure is to calculate the
 speed $\dot Q=\frac {dQ}{dt}=\frac {\partial Q(x)}{\partial x} f(x,u,t)$.
 The second step is to evaluate the gradient of the speed
 $\nabla_u\dot Q$ with respect to  input vector $u$ (speed-gradient vector).
 Finally the law of dynamics is formed as the
 feedback law in the finite form
\begin{equation}\label{22}
 u=-\gamma\nabla_u\dot Q(x,u).
  \end{equation}
 or in the differential form
 \begin{equation} \label{24}
\frac{du}{dt}=-\gamma\nabla_u\dot Q(x,u), \end{equation} where
$\gamma>0$ is a scalar or symmetric matrix {\it gain} (positivity
of a matrix is understood as positive definiteness of associated
quadratic form).
The underlying idea of the choices (\ref{22}) or (\ref{24}) is
that the motion along the antigradient of the speed $\dot Q$
provides decrease of $\dot Q$. It may eventually lead to
negativity of $\dot Q$ which, in turn, yields decrease of $Q$.
Now the speed-gradient principle can be formulated as follows.

\vspace*{0.3cm} {\bf Speed-gradient principle:}{\it~ Among all
possible motions of the system only those are realized for which
the input variables change proportionally  to the speed gradient
$\nabla_u\dot Q(x,u)$ of an appropriate goal functional $Q(x)$. If
there are constraints imposed on the system motion, then the
speed-gradient vector should be projected onto the set of
admissible (compatible with constraints) directions.}
\vspace*{0.3cm}

According to the SG-principle, to describe a system dynamics one
needs to introduce the goal function $Q(x)$. The choice of $Q(x)$
should reflect the tendency of natural behavior to decrease the
current value $Q(x(t))$. Systems obeying the SG-principle will be
called {\it SG-systems}. Below only the  models (\ref{a21}) in a
special form are considered:
\begin{equation}
\label{a22} \dot x=u,
\end{equation}
i.e. a law of change of the state velocities is sought.

Note that the SG-direction is the direction of maximum growth for
$\dot Q(x,u,t)$, i.e.  direction of maximum production rate for
$Q$. Respectively, the opposite direction corresponds to minimum
production rate for $Q$. The finite form (\ref{22}) may be used to
describe irreversible processes, while differential form
(\ref{24}) corresponds to reversible ones. The SG-laws with
nondiagonal gain matrices $\gamma$ can be incorporated if a
non-Euclidean metric in the space of inputs is introduced by the
matrix $\gamma^{-1}$. The matrix $\gamma$ can be used to describe
spatial anisotropy. Admitting dependence of the  matrix $\gamma$
on $x$ one can recover dynamics law for complex mechanical systems
described by Lagrangian or Hamiltonian formalism.
 The SG-principle applies to spatially distributed systems where
  the state $x(t)$ is an element of an infinite dimensional space
 and allows one to model dynamics of spatial fields \cite{Fra90}.

 Consider a simple illustrating example:
motion of a particle in the potential field. In this case the
vector $x={\rm col}\,(x_1,x_2,x_3)$ consists of coordinates
$x_1,x_2,x_3$ of a particle. Choose smooth $Q(x)$ as the potential
energy of a particle and derive the speed-gradient law in the
differential form. To this end, calculate the speed  $ \dot
Q=\left[\nabla_xQ(x)\right]^{\scriptscriptstyle T}u$ and the
speed-gradient $\nabla_u\dot Q=\nabla_xQ(x). $ Then, choosing
differential SG-law (\ref{24}) with the gain $\gamma=m^{-1}$,
where $m>0$ is a parameter, we arrive at familiar Newton's law
$\dot u=-m^{-1}\nabla_xQ(x)$ or $m\ddot x=-\nabla_xQ(x).$


\section{Generalized Onsager relations}

Consider an isolated physical system whose state is characterized
by a set of variables (thermodynamic parameters)
$\xi_1,\xi_2,\ldots,\xi_n$. Let $x_i=\xi_i-\xi_i^*$ be deviations
of the variables from their equilibrium values
$\xi_1^*,\xi_2^*,\ldots,\xi_n^*$. Let the dynamics of the vector
$x_1,x_2,\ldots,x_n$ be described by the differential equations
\begin{equation}
\label{a29} \dot x_i=u_i(x_1,x_2,\ldots,x_n),\quad i=1,2,\ldots,n.
\end{equation}

Linearize equations (\ref{a29}) near equilibrium
\begin{equation}
\label{a30} \dot x_i=-\sum\limits_{k=1}^n\lambda_{ik}x_k,\quad
i=1,2,\ldots,n.
\end{equation}

 The {\it Onsager's principle}
\cite{GP71} claims that the values $\lambda_{ik}$ (kinetic
coefficients) satisfy the equations
\begin{equation}
\label{a31} \lambda_{ik}=\lambda_{ki},\quad i,k=1,2,\ldots,n.
\end{equation}
In general, the Onsager principle is not valid for all systems
e.g. for systems far from  equilibrium. Its existing proofs
 \cite{LL80} require additional postulates. Below a simple new
proof is given, showing that it is valid for irreversible
speed-gradient systems without exceptions.

First of all,  the classical formulation of the Onsager principle
(\ref{a31}) should be extended to nonlinear systems. A natural
extension is the following set of identities:
\begin{equation}
\label{a32} \frac{\partial u_i}{\partial
x_k}(x_1,x_2,\ldots,x_n)=\frac{\partial u_k}{\partial
x_i}(x_1,x_2,\ldots,x_n).
\end{equation}
Obviously, for the case when the system equations (\ref{a29}) have
linear form (\ref{a30}) the identities (\ref{a32}) coincide with
(\ref{a31}). However, since linearization is not used in the
formulation (\ref{a32}) there is a hope that the extended version
of the Onsager law holds for some nonlinear systems far from
equilibrium. The following theorem specifies a class of systems
for which this hope comes true.

\vspace*{0.5cm} {\bf Theorem 1.} {\it There exists a smooth
function $Q(x)$ such that equations (\ref{a29}) represent the
speed-gradient law in finite form for the goal function $Q(x)$ if
and only if the identities (\ref{a32}) hold for all
$x_1,x_2,\ldots,x_n$. }

The proof of the
theorem is very simple. Since (\ref{a29}) is the speed-gradient
law for $Q(x)$, its right-hand sides can be represented in the
form
$
u_i=-\gamma\frac{\partial\dot Q}{\partial u_i},\quad
i=1,2,\ldots,n.
$
Therefore $u_i=-\gamma(\partial Q/\partial x_i)$ (in view of $\dot
Q=(\nabla_xQ)^{\scriptscriptstyle T}u)$. Hence
$
\frac{\partial u_i}{\partial
x_k}=-\gamma\frac{\partial^2Q}{\partial x_i\partial
x_k}=\frac{\partial u_k}{\partial x_i},
$
and identities (\ref{a32}) are valid.
Finally, the condition (\ref{a32}) is necessary and sufficient for
potentiality of the vector-field of the right-hand sides of
(\ref{a29}), i.e. for existence of a scalar function $\bar Q$ such
that $u_i=\gamma\nabla_x\bar Q=\gamma\nabla_u\dot{\bar Q}$.

Thus, for SG-systems the extended form of the Onsager equations
(\ref{a32}) hold without linearization, i.e., they are valid not
only near the equilibrium state. In a special case the condition
(\ref{a32}) was proposed in \cite{Farkas71}. The theorem means
that generalized Onsager relations (\ref{a32}) are necessary and
sufficient for the thermodynamics system to obey the SG-principle
for some $\bar Q$. On the other hand, it is known that different
potential functions for the same potential vector-field can differ
only by a constant: $\bar Q=Q+{\rm const}$ and their stationary
sets coincide. Therefore, if the system tends to maximize its
entropy and the entropy serves as the goal function for the
SG-evolution law, then at every time instant the direction of
change of parameters coincides with the direction maximizing the
rate of entropy change (gradient of the entropy rate). It follows from
Zigler's version of maximum entropy principle \cite{Martyushev06}
that at every time instant it tends to minimize its entropy
production rate (Prigogine principle). That is, if Prigogine
principle holds then the generalized Onsager principle (\ref{a32})
holds and vice versa. Note that for special case the relation
between Prigogine principle and Onsager principle was established
by D.Gyarmati \cite{Gyarmati70}.

For the SG-systems some other properties can be established. Let
for example a system is governed by SG-law with a convex entropy
goal function $S$. Then the decrease of the entropy production
$\dot S$ readily follows from the identities $ \ddot S=d\dot
S/dt=(\nabla_x\dot S)^{\trn}\dot x=
\gamma(\nabla_x||\nabla_xS||^2)^{\trn}\nabla_xS=
2\gamma(\nabla_xS)^{\trn}[\nabla^2_xS](\nabla_xS). $

If the entropy $S(x)$ is convex then its Hessian matrix
$\nabla^2_xS$ is negative semidefinite: $\nabla^2_xS\le 0$. Hence
$\ddot S(x)\le 0$ and $\dot S$ cannot increase \cite{Fra90}.

\section{Speed-gradient entropy maximization} \index{entropy}

It is worth noticing that the speed-gradient principle provides an
answer to the question: {\bf how} the system will evolve? It
differs from the principles of maximum entropy, maximum Fisher
information, etc. providing and answer to the questions: {\bf
where?} and {\bf how far?} Particularly, it means that
SG-principle generates equations for the {\it transient
(nonstationary) mode} rather than the equations for the {\it
steady-state mode} of the system. It allows one to study
nonequilibrium and nonstationary situations, stability of the
transient modes, maximum deviations from the limit mode, etc. Let
us illustrate this feature by example of entropy maximization
problem.

According  to the 2nd thermodynamics law and to the Maximum
Entropy Principle of Gibbs-Jaynes the entropy of any physical
system tends to increase until it achieves its maximum value under
constraints imposed by other physical laws. Such a statement
provides knowledge about the final distribution of the system
states, i.e. about asymptotic behavior of the system when
$t\to\infty$. However it does not provide information about the
way how the system moves to achieve its limit (steady) state.

In order to provide motion equations for the transient mode employ
the SG-principle. Assume for simplicity that the system
 consists of $N$ identical particles distributed over $m$
cells. Let $N_i$ be the number of particles in the $i$th cell and
the mass conservation law holds:
\begin{equation}
\label{sg-ent1} \sum^m_{i=1} N_i=N.
\end{equation}

Assume that the particles can move from one cell to another and we
are interested in the system behavior both in the steady-state and
in the transient modes. The answer for the steady-state case is
given by the Maximum Entropy Principle: if nothing else is known
about the system, then its limit behavior will maximize its
entropy \cite{Jaynes57}. Let the entropy of the system be defined
as logarithm of the number of possible states:
\begin{equation}
\label{sg-ent2} S=\ln\frac{N!}{N_1!\cdot\dots\cdot N_m!}.
\end{equation}
If there are no other constraints except normalization condition
(\ref{sg-ent1}) it achieves maximum when $N^*_i=N/m$. For large
$N$ an approximate expression is of use. Namely, if the number of
particles $N$ is large enough, one may use the Stirling
approximation $N_i!\approx (N_i/e)^N.$ Then
$$
S\approx N\ln\frac{N}{e}-\sum^m_{i=1} N_i\ln\frac{N_i}{e}=
-\sum^m_{i=1} N_i\ln\frac{N_i}{N}
$$
 which coincides with the standard definition for the entropy $S=-\sum^m_{i=1} p_i\ln
 p_i$, modulo a constant multiplier $N$,
 if the probabilities $p_i$ are understood as frequencies $N_i/N$.

To get an  answer for transient mode  apply the SG-principle
choosing the entropy $S(X)=-\sum^m_{i=1} N_i\ln N_i$ as the goal
function to be maximized, where $X={\rm col}(N_1,\dots, N_m)$ is
the state vector of the system.
Assume for simplicity
that the motion is continuous in time and the numbers $N_i$ are
changing continuously, i.e. $N_i$ are not necessarily integer (for
large $N_i$ it is not a strong restriction). Then the sought law
of motion can be represented in the form
\begin{equation}
\label{sg-ent3} \dot N_i=u_i,~i=1,\dots ,m,
\end{equation}
where $u_i=u_i(t),~i=1,\dots,m$ are controls -- auxiliary
functions to be determined. According to the SG-principle one
needs to evaluate first the speed of change of the entropy
(\ref{sg-ent2}) with respect to the system (\ref{sg-ent3}), then
evaluate the gradient of the speed with respect to the vector of
controls $u_i$ considered as frozen parameters and finally define
actual controls proportionally to the projection of the
speed-gradient to the surface of constraints (\ref{sg-ent1}). In
our case the goal function is the entropy $S$ and its speed
coincides with the entropy production $\dot S$. In order to
evaluate $\dot S$ let us again approximate $S$ from the Stirling
formula $N_i!\approx (N_i/e)^N$:
\begin{equation}
\label{sg-ent4} \hat S=N \ln N-N-\sum^m_{i=1} (N_i\ln
N_i-N_i)=N\ln N - \sum^m_{i=1} N_i\ln N_i.
\end{equation}
Evaluation of $\dot{\hat S}$ yields $$\dot{\hat S}=-\sum^m_{i=1}
((u_i\ln N_i+N_i\frac{u_i}{N_i})=-\sum^m_{i=1} u_i(\ln N_i+1).
$$
It follows from (\ref{sg-ent1}) that $\sum^m_{i=1} u_i=0$. Hence
$\dot{\hat S}=-\sum^m_{i=1} u_i\ln N_i$. Evaluation of the
speed-gradient yields $\frac{\partial\dot{\hat S}}{\partial
u_i}=-\ln N_i$ and the  SG-law
$\label{sg-ent5} u_i=\gamma(-\ln N_i+\lambda),~i=1,\dots ,m,$
where Lagrange multiplier $\lambda$ is chosen in order to fulfill
the constraint $\sum^m_{i=1} u_i=0$, i.e.
$\lambda=\frac{1}{m}\sum^m_{i=1} \ln N_i$. The final form of the
system dynamics law is as follows:
\begin{equation}
\label{sg-ent6} \dot N_i=\frac{\gamma}{m}\sum^m_{i=1} \ln
N_i-\gamma\ln N_i,~i=1,\dots ,m.
\end{equation}

According to the SG-principle the equation  (\ref{sg-ent6})
determines transient dynamics of the system.  To confirm
consistency of the choice (\ref{sg-ent6}) let us find the
steady-state mode, i.e. evaluate asymptotic behavior of the
variables $N_i$. To this end note that in the steady-state $\dot
N_i=0$ and $\sum^m_{i=1} \ln N_i=\ln N_i$. Hence all $N_i$ are
equal: $N_i=N/m$ which corresponds to the maximum entropy state
and agrees with thermodynamics.

The next step is to examine  stability of the steady-state mode.
It can be done by means of the entropy Lyapunov function
$ V(X)=S_{max}-S(X)\ge 0,$
where $S_{max}=N\ln m$. Evaluation of $\dot V$ yields $$\dot
V=-\dot S=\sum^m_{i=1}u_i\ln
N_i=\frac{\gamma}{m}\big[(\sum^m_{i=1} \ln
N_i)^2-m\sum^m_{i=1}(\ln N_i)^2\big].$$ It follows from the
Cauchy-Bunyakovsky-Schwarz inequality that $\dot V(X)\le 0$ and
the equality $\dot V(X)=0$ holds if and only if all the values
$N_i$ are equal, i.e. only at the maximum entropy state. Thus the
law (\ref{sg-ent6}) provides global asymptotic stability of the
maximum entropy state. The physical meaning of the law
(\ref{sg-ent6}) is moving along the direction of the maximum
entropy production rate (direction of the fastest entropy growth).

The case of more than one constraint can be treated in the same
fashion. Let in addition to the mass conservation law
(\ref{sg-ent1}) the
 energy conservation law hold.  Let $E_i$ be the energy of the particle
in the $i$th cell and the total energy $E=\sum^m_{i=1}N_iE_i$ be
conserved. The energy conservation law
\begin{equation}
\label{sg-ent8}E=\sum^m_{i=1}N_iE_i
\end{equation}
appears as an additional constraint. Acting in a similar way, we
arrive at the law (\ref{sg-ent6}) which needs modification to
ensure conservation of the energy (\ref{sg-ent8}). According to
the SG-principle one should form the projection onto the  surface
(in our case -- subspace of dimension $m-2$) defined by the
relations
\begin{equation}
\label{sg-ent9}\sum^m_{i=1}u_iE_i=0, ~~\sum^m_{i=1}u_i=0.
\end{equation}
It means that the evolution law should have the form
\begin{equation}
\label{sg-ent10} u_i=\gamma(-\ln
N_i)+\lambda_1E_i+\lambda_2,~i=1,\dots ,m,
\end{equation}
where $\lambda_1, \lambda_2$ are determined by substitution of
(\ref{sg-ent10}) into (\ref{sg-ent9}). The obtained equations are
linear in  $\lambda_1, \lambda_2$ and their solution is given by
formulas
\begin{equation}
\begin{cases}
\label{sg-ent10a}  \lambda_1=\frac{\gamma m\sum^m_{i=1}E_i\ln
N_i)-\gamma (\sum^m_{i=1}E_i)(\sum^m_{i=1}\ln
N_i)}{m\sum^m_{i=1}E_i^2-(\sum^m_{i=1}E_i)^2},\cr \cr
 \lambda_2=\frac{\gamma}{m}\sum^m_{i=1}\ln
 N_i-\frac{\lambda_1}{m}\sum^m_{i=1}E_i.
\end{cases}
\end{equation}
The solution of (\ref{sg-ent10a}) is well defined if
$m\sum^m_{i=1}E_i^2-(\sum^m_{i=1}E_i)^2\ne 0$ which holds unless
all the $E_i$ are equal (degenerate case).

Let us evaluate the equilibrium point of the system
(\ref{sg-ent3}), (\ref{sg-ent10}) and  analyze its stability. At
the equilibrium point of the system the following equalities
hold: $ \gamma(-\ln N_i)+\lambda_1E_i+\lambda_2=0,~i=1,\dots ,m.$
Hence
\begin{equation}
\label{sg-ent11}
 N_i= C\exp(-\mu E_i),~i=1,\dots ,m,
\end{equation}
 where $\mu=\lambda_1/\gamma$ and $C=\exp(-\lambda_2/\gamma).$

The value of $C$ can also be chosen from the normalization
condition $C=N(\sum^m_{i=1}\exp{(-\mu E_i)})$. We see that
equilibrium of the system with conserved energy corresponds to the
Gibbs distribution which agrees with classical thermodynamics.
Again it is worth to note that the direction of change of the
numbers $N_i$ coincides with the direction of the fastest growth
of the local entropy production subject to constraints. As before,
it can be shown that $ V(X)=S_{max}-S(X)$ is Lyapunov function for
the system and that the Gibbs distribution is the only stable
equilibrium of the system in nongenerate cases. Similar results
are valid for continuous (distributed) systems even for more
general problem of minimization of relative entropy (Kullback
divergence) \cite{F07}.

\section*{Conclusions}

Speed-gradient variational principle provides a useful yet simple
 addition to
classical results in thermodynamics. Whereas the classical results
allow researcher to answer the question ``Where it goes to?'', the
speed-gradient approach provides an answer to the question: ``How
it goes and how it reaches  its steady-state mode?'' SG-principle may be
applied to evaluation of nonequilibrium stationary states and
study of system internal structure evolution \cite{Khantuleva05},
description of transient dynamics of complex networks
 \cite{Rangan06,Fronczak06}, etc.
A different approach to variational description of nonstationary
nonequilibrium processes is proposed in \cite{Osipov93}.

\medskip

The work was supported by Russian Foundation for Basic Research
(project RFBR 05-01-00869).

\end{document}